\pdfoutput=1

\documentclass[3p,times,twocolumn]{elsarticle}

\usepackage{ecrc}

\volume{00}

\firstpage{1}

\journalname{Nuclear Physics B Proceedings Supplement}

\runauth{B. Sarrazin et al.}

\jid{nuphbp}

\jnltitlelogo{Nuclear Physics B Proceedings Supplement}

\usepackage{amssymb}

\usepackage[figuresright]{rotating}

\usepackage{hyperref}
\hypersetup{breaklinks=true}
\usepackage{amsmath}
\DeclareMathOperator{\sgn}{sgn}

\newcommand{\bsmm}{\mathcal{B}(B_s\to\mu\mu)}
\newcommand{\bsgamma}{\mathcal{B}(b\to s\gamma)}
\newcommand{\btaunu}{\mathcal{B}(B\to\tau\nu)}

\begin{document}

\begin{frontmatter}



\dochead{}

\title{How alive is constrained SUSY really?}


\author[addr:bonn]{Philip Bechtle}
\ead{bechtle@physik.uni-bonn.de}
\author[addr:bonn]{Klaus Desch}
\ead{desch@physik.uni-bonn.de}
\author[addr:bonn,addr:bethe]{Herbert K. Dreiner}\ead{dreiner@th.physik.uni-bonn.de}
\author[addr:rio]{Matthias Hamer}
\ead{mhamer@cbpf.br}
\author[addr:aachen,addr:slac]{Michael Kr\"amer}
\ead{mkraemer@physik.rwth-aachen.de}
\author[addr:wuerzburg]{Ben O'Leary}
\ead{ben.oleary@physik.uni-wuerzburg.de}
\author[addr:wuerzburg]{Werner Porod}
\ead{porod@physik.uni-wuerzburg.de}
\author[addr:bonn]{Bj\"orn Sarrazin\corref{speaker}}
\ead{sarrazin@physik.uni-bonn.de}
\author[addr:santacruz]{Tim Stefaniak}
\ead{tistefan@ucsc.edu}
\author[addr:bonn]{Mathias Uhlenbrock}
\ead{uhlenbrock@physik.uni-bonn.de}
\author[addr:bonn]{Peter Wienemann}
\ead{wienemann@physik.uni-bonn.de}

\address[addr:bonn]{Physikalisches Institut, University of Bonn, Germany}
\address[addr:bethe]{Bethe Center for Theoretical Physics, University of Bonn, Germany}
\address[addr:rio]{Centro Brasileiro de Pesquisas Fisicas, Rio de Janeiro, Brazil}
\address[addr:aachen]{Institute for Theoretical Particle Physics and Cosmology, RWTH Aachen, Germany}
\address[addr:slac]{SLAC National Accelerator Laboratory, Stanford University, USA}
\address[addr:wuerzburg]{Institut f\"ur Theoretische Physik und Astrophysik, University of W\"urzburg, Germany}
\address[addr:santacruz]{Santa Cruz Institute for Particle Physics, University of California, Santa Cruz, USA}
\cortext[speaker]{Speaker}

\begin{abstract}
Constrained supersymmetric models like the CMSSM might look less attractive nowadays because of fine tuning arguments. They also might look less probable in terms of Bayesian statistics. The question how well the model under study describes the data, however, is answered by frequentist $p$-values. Thus, for the first time, we calculate a $p$-value for a supersymmetric model by performing dedicated global toy fits. We combine constraints from low-energy and astrophysical observables, Higgs boson mass and rate measurements as well as the non-observation of new physics in searches for supersymmetry at the LHC. Using the framework \textsc{Fittino}, we perform global fits of the CMSSM to the toy data and find that this model is excluded at more than $95\%$ confidence level.
\end{abstract}


\end{frontmatter}



\section{Introduction}\label{sec:introduction} 

As a fully renormalizable theory beyond the Standard Model (SM) supersymmetry \cite{Wess:1974tw} makes a wide range of predictions, allowing to probe it with many different measurements.
Indirect constraints from low energy measurements, the discovery of an SM-like Higgs boson, direct searches for supersymmetric particles and additional Higgs bosons as well as astrophysical observations constrain the parameter space. 

As one of the simplest supersymmetric models the Constrained Minimal Supersymmetric Standard Model (CMSSM) \cite{Nilles:1983ge} has been studied extensively in light of this data over the last years. 
It has five free parameters:
The ratio of the vacuum expectation values of the two Higgs doublets ($\tan\beta$), the sign of the Higgs mixing parameter ($\sgn\mu$), and universal soft-breaking parameters defined at the GUT scale for scalar masses ($M_0$), gaugino masses ($M_{1/2}$), and trilinear couplings ($A_0$). 
Global fits are used to determine the parameter region which is in best agreement with these measurements. 

It has become clear from these studies that the non-observation of convincing hints of new physics at the LHC becomes challenging for this model once constraints from low energy observables are taken into account. However, it has so far been an open question what this means quantitatively for the validity of the model.
In this work, we address this question for the first time systematically using global toy fits to calculate a $p$-value for the CMSSM.

The remainder of this article is organized as follows: Section~\ref{sec:status} gives a brief review of the status of the CMSSM.
Section~\ref{sec:pvalue} introduces the $p$-value.
Section~\ref{sec:measurements} details which measurements we use for our analysis, while Section~\ref{sec:predictions} summarizes the codes used to obtain the corresponding model predictions.
The technique to compare the measurements and their predictions is described in Section~\ref{sec:techniques}.
Profile likelihood based results are shown in Section~\ref{sec:pl}, while results from toy fits are given in Section~\ref{sec:toys}.
We conclude in Section~\ref{sec:conclusions}. 

\section{What has happened so far}\label{sec:status} 

Before the start of the LHC, global fits of the CMSSM preferred relatively low values of the mass parameters.
This was mainly driven by the anomalous magnetic moment of the muon $a_{\mu}$ and one of the reasons why it was anticipated to find supersymmetry early at the LHC.
Since this has not happened yet, some tension arises between low energy observables and exclusions from direct searches.
While still giving an excellent fit, this effect could already be seen after the first searches for supersymmetry by ATLAS or CMS with an integrated luminosity of $35$\,pb$^{-1}$ at $7$\,TeV were included in the fit \cite{Bechtle:2011dm}.
With increasing luminosity up to $20\,\mathrm{fb}^{-1}$ at 8\,TeV this tension increased significantly \cite{Bechtle:2012zk, Bechtle:2013mda}.

While there is also no convincing hint for new physics from direct and indirect astrophysical searches for dark matter, bounds from these experiments are currently easily fulfilled by the CMSSM.
Similarly, while no extended Higgs sector has been found, the discovery of an SM-like Higgs boson is in agreement with CMSSM predictions.
The inclusion of numerous SM-like Higgs rate measurements even improves the $\chi^2/\mathrm{ndf}$ ratio of the fit \cite{Bechtle:2013mda}.
This is due to the fact that the CMSSM, when forced to large supersymmetric masses by direct searches, enters the so-called decoupling limit where an SM-like Higgs is predicted.

\section{The missing piece}\label{sec:pvalue} 

The status described in Section~\ref{sec:status} seems to leave us in an unsatisfactory situation after the end of the first run of the LHC.
On the one hand, there is some tension between low energy measurements and direct searches, on the other hand, the model describes many measurements very well and there is no single clear piece of evidence against it.
This hampers a final judgement.
Here, we escape from this dead end by calculating the $p$-value of the CMSSM.
In high energy physics, the $p$-value is the standard measure for the agreement between model and data. 
It is defined as the probability to obtain an agreement at least as bad as the one observed assuming the model is true.

Determining the $p$-value in global fits is challenging because the $\chi^2$-function used in the fit is not necessarily $\chi^2$-distributed due to non-Gaussian uncertainties and non-linear dependences of the observables on the model parameters.
Therefore, it is not possible to calculate the $p$-value by simply integrating over the $\chi^2$-distribution for the appropriate number of degrees of freedom.
Instead, it is necessary to determine the shape of the distribution by performing toy fits as will be described in Section~\ref{sec:techniques}.
Using the framework \textsc{Fittino} \cite{Bechtle:2004pc}, this makes it possible to calculate a confidence level for an exclusion of the CMSSM for the first time.

\section{Measurements}\label{sec:measurements} 

We use the same set of low energy observables as in Ref.~\cite{Bechtle:2012zk}, but with updated measurements as given in Table \ref{tab:measurements}.
They include the anomalous magnetic moment of the muon $a_\mu$, the effective weak mixing angle $\sin^2 \theta_\mathrm{eff}$, the masses of the top quark and $W$ boson, the oscillation frequency $\Delta m_s$, as well as the branching ratios $\bsmm$, $\btaunu$, and $\bsgamma$. 

In order to have a good dark matter candidate, we require the lightest supersymmetric particle to be a neutralino.
We use the dark matter relic density $\Omega h^2=0.1187 \pm 0.0017$ as measured by the Planck collaboration \cite{Ade:2013zuv} and bounds on the spin-independent WIMP-nucleon scattering cross section as measured by the LUX experiment \cite{Akerib:2013tjd}. 

Concerning direct searches at the LHC, we use the final state with jets and  missing transverse momentum \cite{Aad:2014wea} because this channel is by far the most dominant one in the region preferred by the low energy observables. In addition, we make sure that the LEP bound on the chargino mass, $m_{\tilde{\chi}^\pm_1}>103.5$\,GeV, is fulfilled \cite{LEPSUSYWG}.
We apply the limits from Higgs boson searches as implemented in \textsc{HiggsBounds 4.1.1} \cite{Bechtle:2013wla}. Using the program \textsc{HiggsSignals 1.2.0}, we include Higgs signal rate measurements in the channels $h\to \gamma\gamma$, $h\to ZZ$, $h\to W W$,  $Vh\to Vbb$, and $h\to \tau\tau$ from both ATLAS and CMS. Higgs boson mass measurements in the channels $h\to \gamma\gamma$ and $h\to ZZ$ are used as well ( see Ref.~\cite{Bechtle:2013xfa} and references therein).
\begin{table}
\begin{tabular}{llc}
\hline\noalign{\smallskip}
      $ a_{\mu}-a_{\mu}^{\mathrm{\mathrm{SM}}}$ & $(28.7 \pm 8.0 )\times10^{-10}$      & \cite{Bennett:2006fi,Davier:2010nc}\\
      $ \sin^2\theta_{\mathrm{eff}}$            & $0.23113 \pm 0.00021 $          & \cite{ALEPH:2005ab}\\
      $m_t$                       & $(173.34 \pm 0.27 \pm 0.71)$\,GeV         & \cite{ATLAS:2014wva}\\                                                     
      $m_W$                                     & $(80.385 \pm 0.015 )$\,GeV       & \cite{Group:2012gb} \\
       $ \Delta m_{s}$                           & $(17.719 \pm  0.036 \pm 0.023 )\,\mathrm{ps}^{-1}$  & \cite{Beringer:1900zz} \\
      $ {\cal B}(B_s\to\mu\mu) $                & $(2.90 \pm 0.70 )\times 10^{-9}$                   & \cite{CMSandLHCbCollaborations:2013pla}\\
      $ {\cal B}(b\to s\gamma)$ & $(3.43 \pm  0.21 \pm 0.07 )\times 10^{-4}$       & \cite{Amhis:2012bh}\\
      $ {\cal B}(B\to\tau\nu)$   & $(1.05 \pm 0.25 ) \times 10^{-4}$                  & \cite{Beringer:1900zz}\\
\noalign{\smallskip}\hline
\end{tabular}
\caption{ Low energy observables used in the fit.}\label{tab:measurements}\end{table}

\section{Model predictions}\label{sec:predictions} 
\begin{table}
\begin{tabular}{lc}
\hline\noalign{\smallskip}
      $ a_{\mu}-a_{\mu}^{\mathrm{\mathrm{SM}}}$   & $7\%$    \\
      $ \sin^2\theta_{\mathrm{eff}}$                 & $0.05\%$    \\
      $m_t$                       & 1\,GeV  \\                                                       
      $m_W$                                     & $0.01\%$  \\
      $ \Delta m_{s}$                          & $24\%$   \\
            $ {\cal B}(B_s\to\mu\mu) $                & $26\%$                     \\
      $ {\cal B}(b\to s\gamma)$  & $14\%$   \\
      $ {\cal B}(B\to\tau\nu)$      & $20\%$        \\    
\noalign{\smallskip}\hline
\end{tabular}
\caption{Theoretical uncertainties of the low energy observables used in the fit.}\label{tab:theounc}
\end{table}

To calculate the corresponding model predictions of the observables, we employ the following public codes:
The spectrum calculation is done using \textsc{SPheno 3.2.4} \cite{Porod:2003um, Porod:2011nf}. Properties of the Higgs bosons as well as $a_{\mu}$,  $\Delta m_s$, $\sin^2\theta_\mathrm{eff}$ and $m_W$ are taken from \textsc{FeynHiggs 2.10.1} \cite{Heinemeyer:1998yj}, which -- compared to \textsc{FeynHiggs 2.9.5} and earlier versions -- contains a significantly improved calculation of the Higgs boson mass \cite{Hahn:2013ria}.
B-physics branching ratios are calculated by \textsc{SuperIso 3.3} \cite{Mahmoudi:2008tp}. 
Concerning astrophysical observables we use \textsc{MicrOMEGAs 3.6.9} \cite{Belanger:2001fz, Belanger:2004yn} for the dark matter relic density and \textsc{DarkSUSY 5.0.5} \cite{Gondolo:2004sc} via the interface program \textsc{AstroFit} \cite{AstroFit} for the direct detection cross section.
For the calculation of expected event numbers at the LHC, we use the Monte Carlo event generator \textsc{Herwig++} \cite{Bahr:2008pv} and a carefully tuned and validated version of the fast parametric detector simulation \textsc{Delphes} \cite{Ovyn:2009tx}.
We reweight the events depending on their production channel according to NLO cross sections obtained from \textsc{Prospino} \cite{Beenakker:1996ed}. Renormalization and factorization scales have been chosen such that NLO+NLL results are reproduced.
For all predictions we take theoretical uncertainties into account, most of them are parameter dependent. For the low energy observables, they are given in Table \ref{tab:theounc}. For the dark matter relic density we assume a theoretical uncertainty of $10\%$, for the neutralino-nucleon cross section $50\%$, for the Higgs boson mass prediction $2.4\%$, and for Higgs rates we use the uncertainties given by the LHC Higgs Cross Section Working Group \cite{Heinemeyer:2013tqa}.

\section{Techniques}\label{sec:techniques} 

The framework \textsc{Fittino} compares measurements and model predictions using a $\chi^2$-function given by
\begin{displaymath}
  \chi^2 = \left(\vec{O}_\mathrm{meas} - \vec{O}_\mathrm{pred}\right)^T \,\mathrm{cov}^{-1} \, \left(\vec{O}_\mathrm{meas}-\vec{O}_\mathrm{pred}\right) + \chi^2_\mathrm{limits},
\end{displaymath}
where $\vec{O}_\mathrm{meas}$ is the vector of measurements, $\vec{O}_{\mathrm{pred}}$ the vector of corresponding model predictions, and $\mathrm{cov}$ the covariance matrix containing all experimental and theo\-retical uncertainties and their correlations. 
The term $\chi^2_\mathrm{limits}$ contains the contributions from Higgs limits, LUX and the direct searches.
This $\chi^2$ function is used in a Markov chain Monte Carlo in order to sample the CMSSM parameter space efficiently.
We vary the continuous CMSSM parameters $M_0$, $M_{1/2}$, $\tan\beta$ and $A_0$ and fix $\sgn\mu=+1$ in order to get positive supersymmetric contributions to the anomalous magnetic moment of the muon. We fit the top quark mass as an additional free parameter. 

In order to determine the $p$-value, we perform toy fits.
As described in section~\ref{sec:pvalue} the $p$-value is given by
\begin{displaymath}
 p=\int_{\chi^2_\mathrm{min,\,obs}}^\infty f(x)\,dx
\end{displaymath} 
where $\chi^2_\mathrm{min,\, obs}$ is the minimal observed $\chi^2$ and  $f(x)\,dx$ is the probability to get a minimal $\chi^2$ between $x$ and $x+dx$ if the model is true.
Toy fits are needed in order to determine the a priori unknown distribution $f(x)$.
We smear the observable values within their uncertainties around their predictions at the best fit point and fit the model to each of these toy measurements.
In the limit of large statistics, the $p$-value is then by definition given as the fraction of toy fits that have a minimal $\chi^2$ at least as bad as the one observed.

\section{Profile likelihood based results}\label{sec:pl} 

\begin{figure}[t] 
\includegraphics[width=0.5\textwidth]{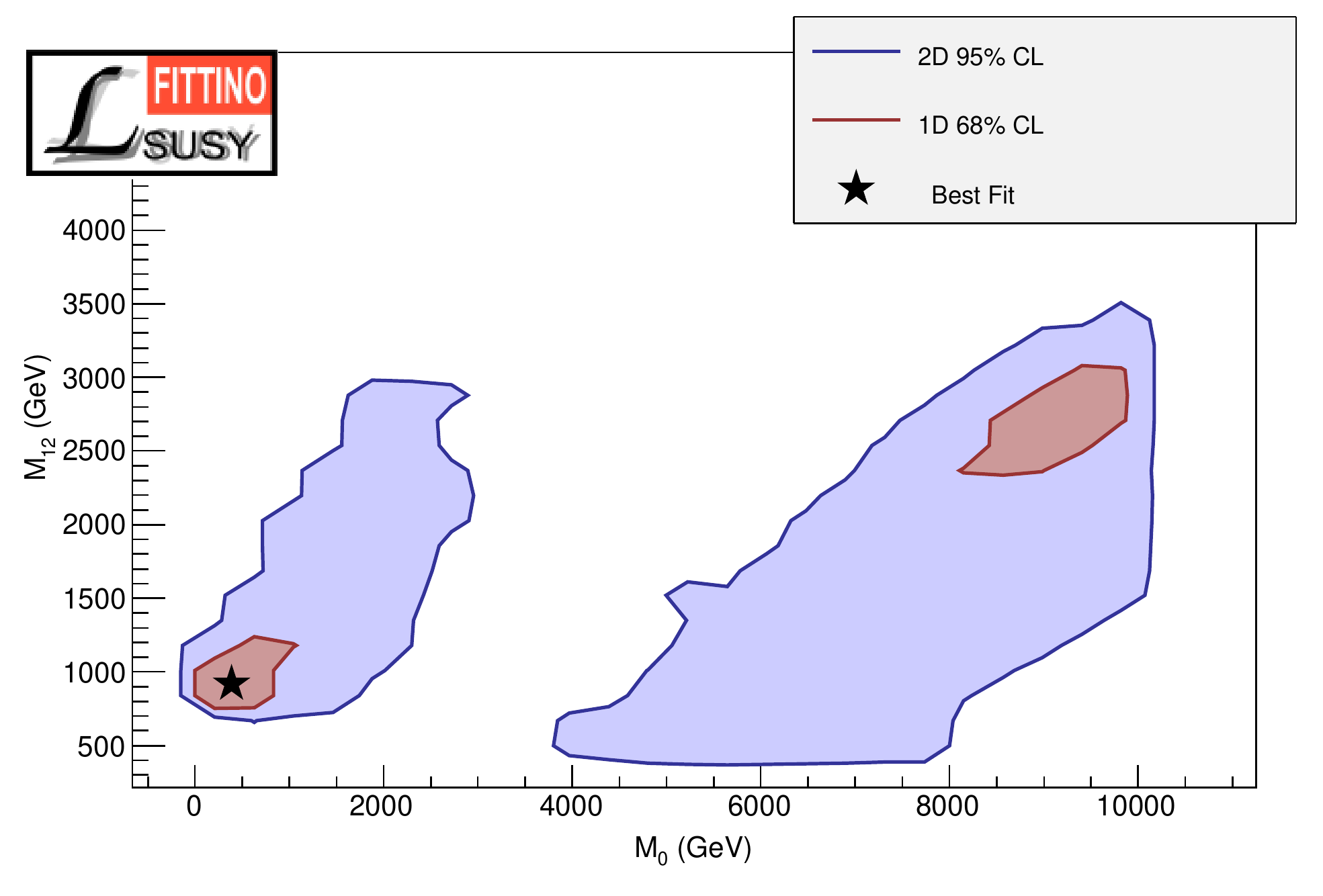}
\caption{Preferred parameter region in the ($M_0$, $M_{1/2}$) plane.}\label{fig:m0m12}
\end{figure}

\begin{figure}[t] 
\includegraphics[width=0.5\textwidth]{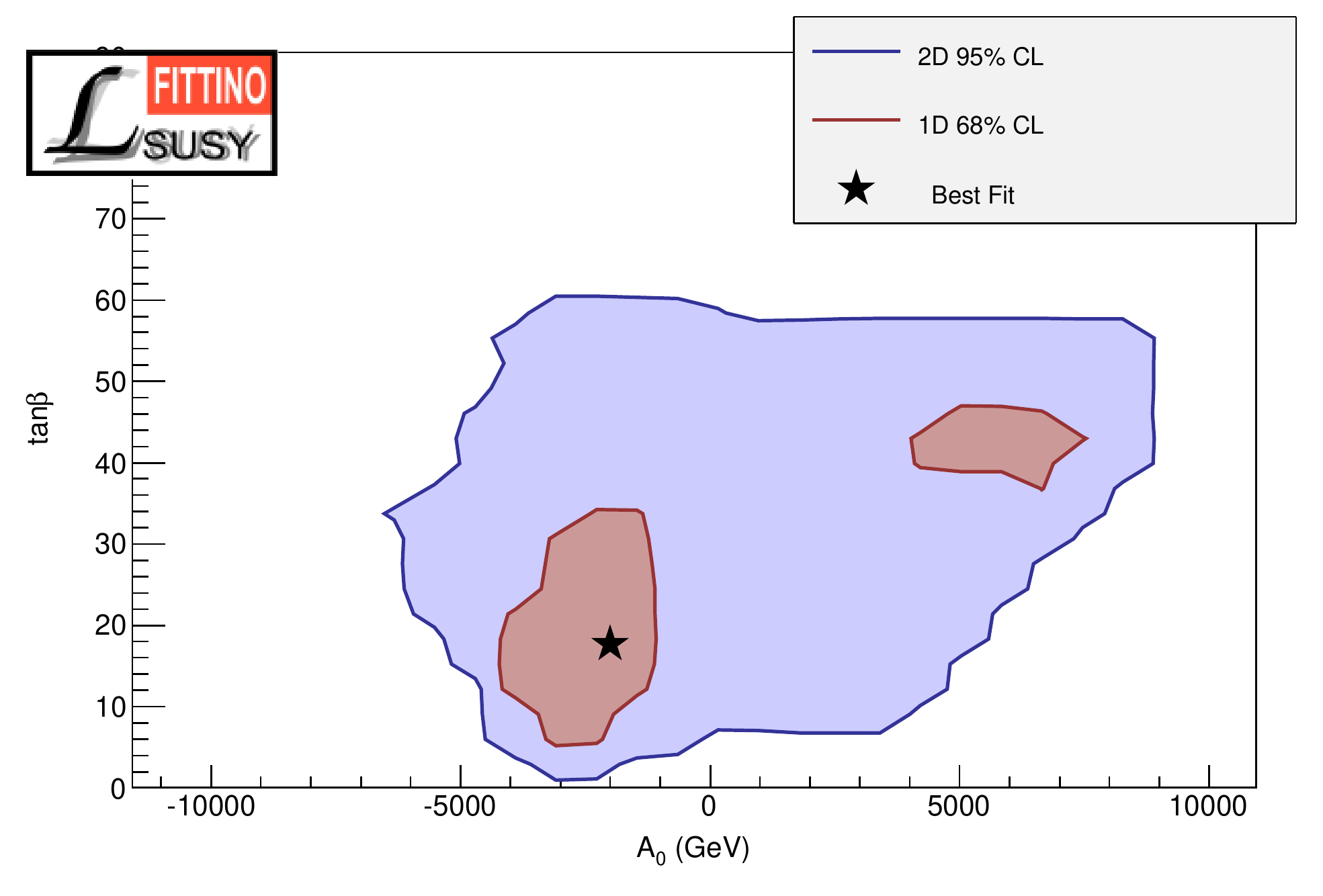}
\caption{Preferred parameter region in the ($A_0$, $\tan\beta$) plane.}\label{fig:a0tanbeta}
\end{figure}

We find a minimal $\chi^2$ of $\chi^2_\mathrm{min, obs}=30.4$ with $22$ degrees of freedom.  
The best fit point is shown in Figures~\ref{fig:m0m12} and \ref{fig:a0tanbeta} as a black star, together with the projections of the $\Delta\chi^2\leq1$ and $\Delta\chi^2\leq6$ regions in red and blue, respectively.
Figure~\ref{fig:m0m12} shows two disjoint areas in the ($M_0$, $M_{1/2}$) plane, which are mainly characterized by different dominant dark matter annihilation mechanisms.
At lower masses, stau coannihilation and the Higgs funnel are present.
At higher masses, the focus point region and chargino coannihilation are relevant.
While the high mass region has been disfavored in our previous fits \cite{Bechtle:2013mda}, it is now even part of the $\Delta\chi^2\leq1$ region.
This is due to the new Higgs boson mass calculation in \textsc{FeynHiggs}, which allows larger Higgs boson masses when going to large supersymmetric mass scales.
We did not scan the parameter space for values of $M_0$ larger than $10$\,TeV. 
Figure~\ref{fig:a0tanbeta} shows the ($A_0$, $\tan\beta$) plane.
In the low mass region negative values of $A_0$ are preferred and a wide range of values of $\tan\beta$ is allowed, whereas the high mass region prefers positive values of $A_0$ and larger values of $\tan\beta$ above 40.

\begin{figure}[hbt]
\includegraphics[width=0.5\textwidth]{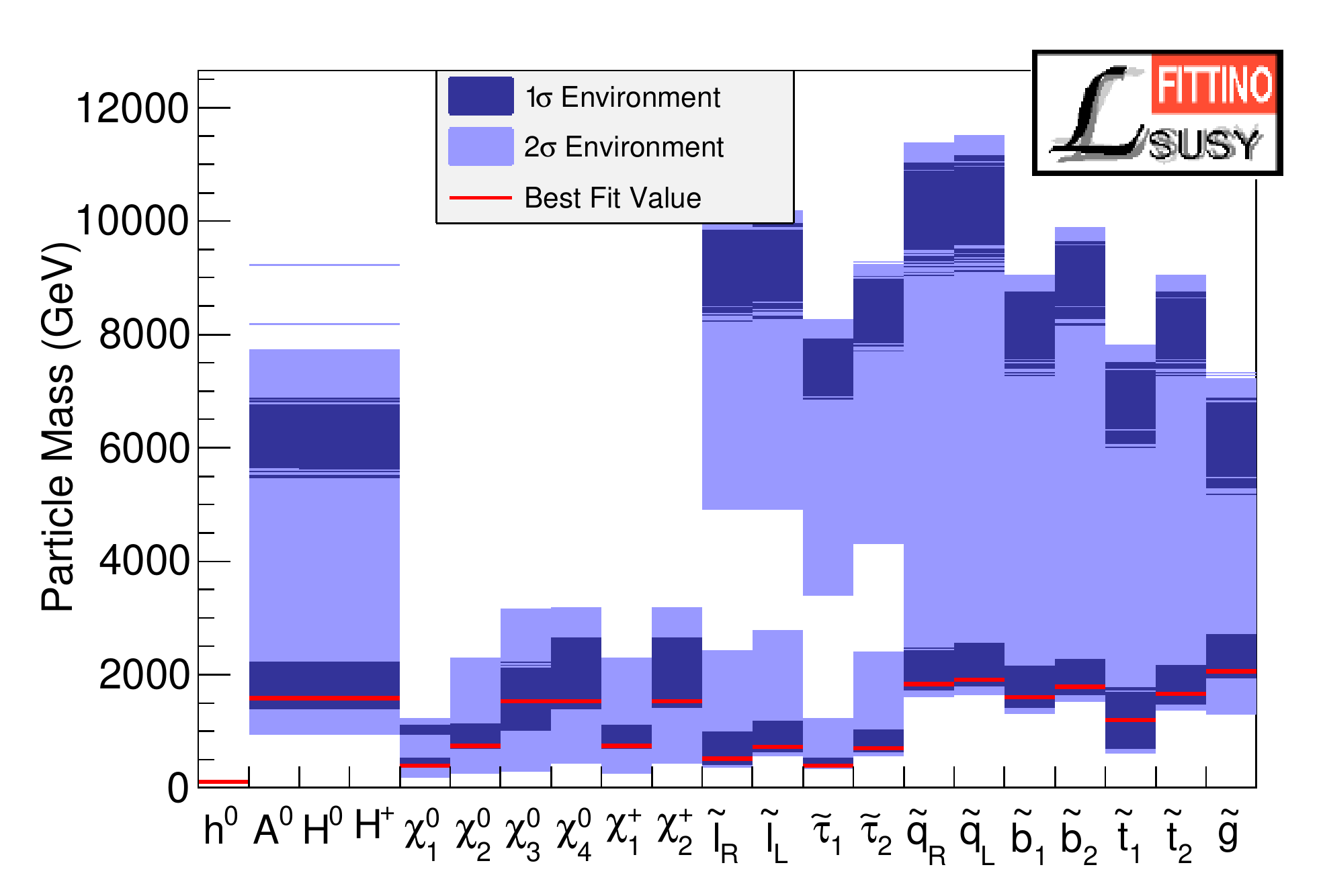}
\caption{Predicted mass spectrum.}\label{fig:mass}
\end{figure}

The corresponding mass spectrum is shown in Figure \ref{fig:mass}. 
Due to the high mass region, squark masses of about 10 TeV are now allowed at $1\sigma$. At the best fit point, squarks and gluinos have masses of about 2\,TeV, the heavy Higgs boson masses are about 1.5\, TeV. Charginos, neutralinos and sleptons stay lighter. 

\begin{figure}[hbt]
\includegraphics[width=0.5\textwidth]{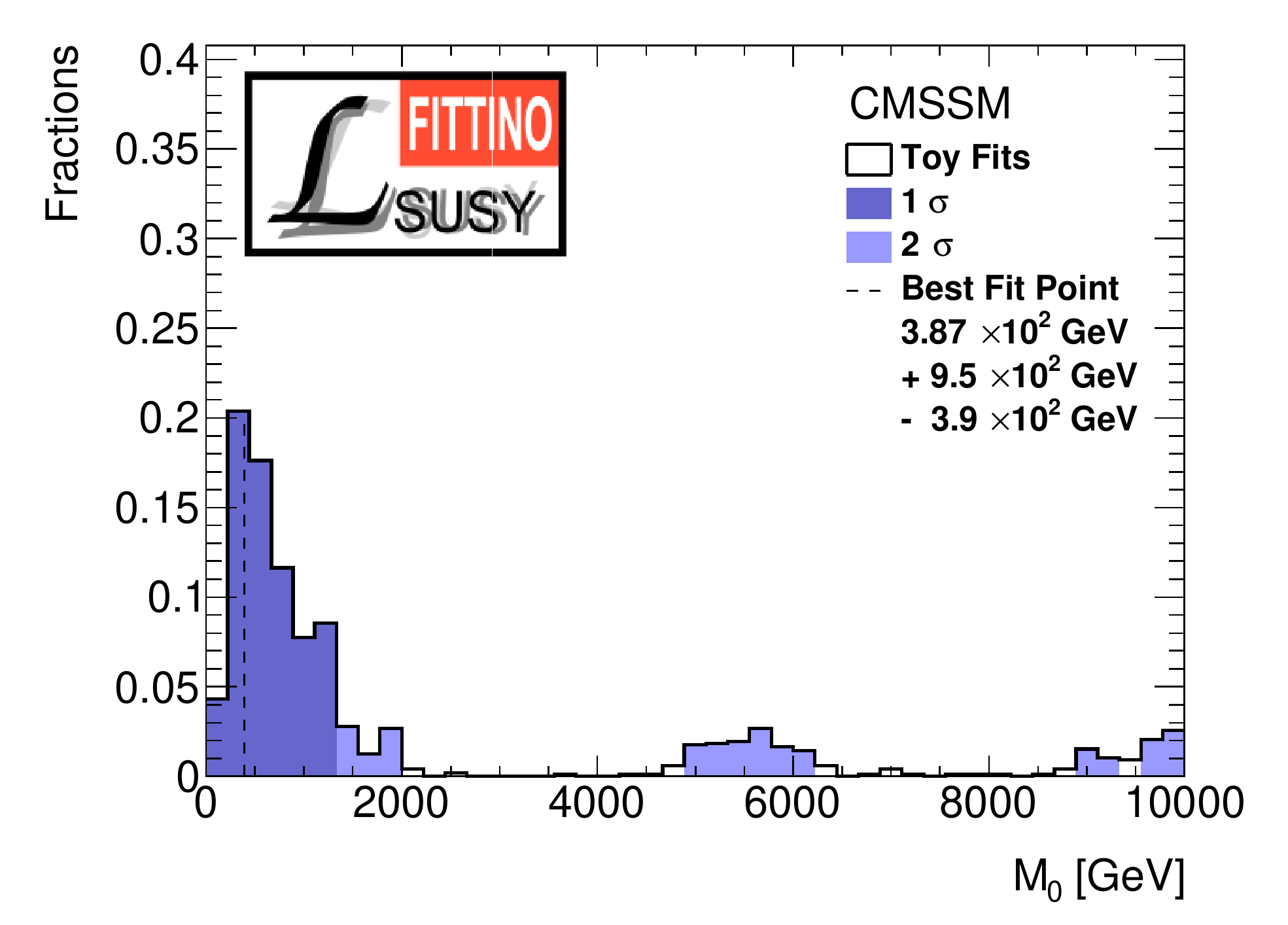}
\caption{ Distribution of best fit values of $M_0$ obtained from toy fits.}\label{fig:m0}
\end{figure}

\section{Toy fit results}\label{sec:toys} 

We repeat the fit described above about 1000 times with toy measurements. The obtained distribution of best fit points is shown in $M_0$ in Figure \ref{fig:m0}. Most toy fits have their minimum in the low mass region, but some of them end up in the high mass region.

\begin{figure}[t]
\includegraphics[width=0.5\textwidth]{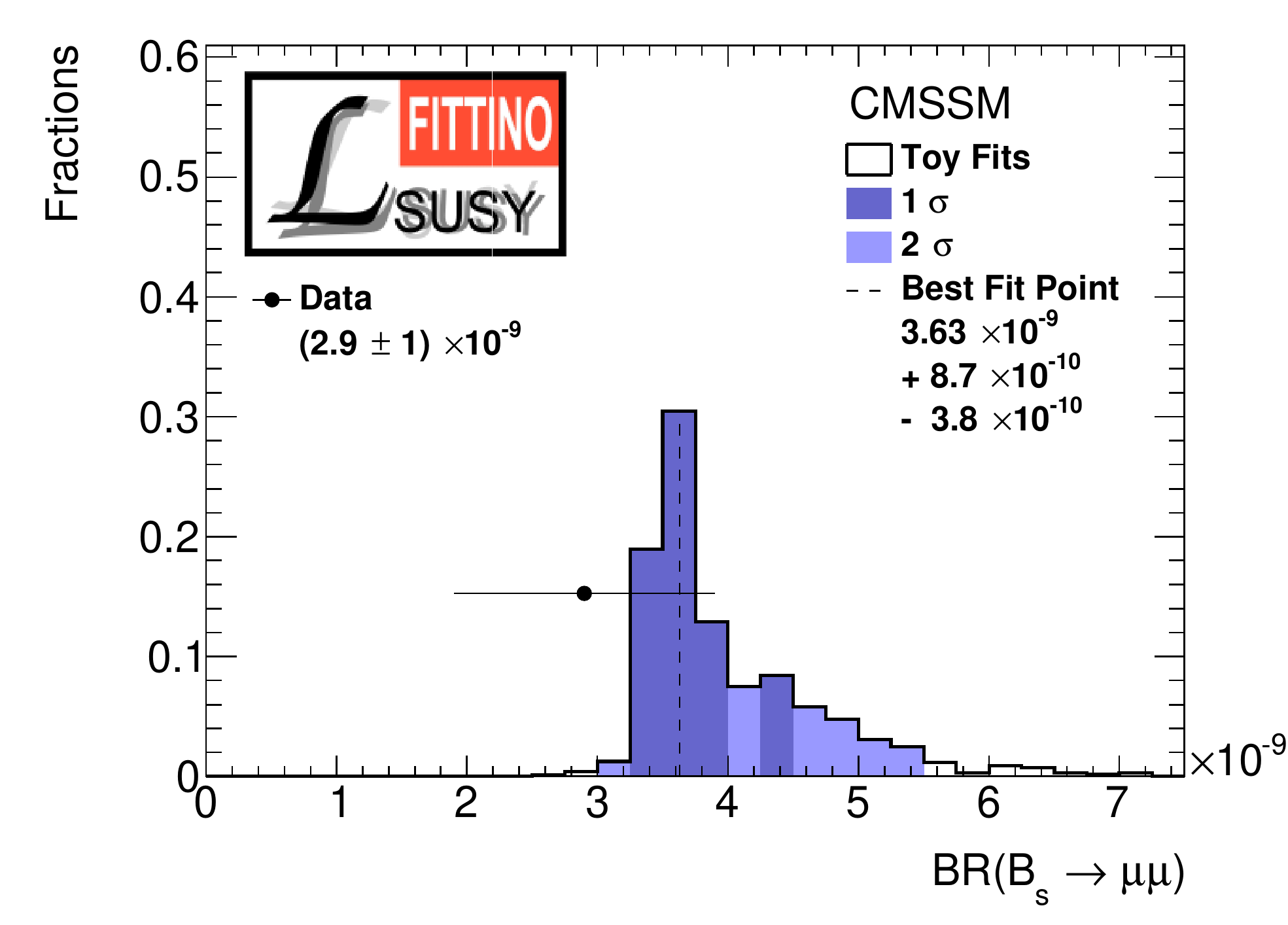}
\caption{ Distribution of predicted values of $\bsmm$ at the best fit points obtained from toy fits.}\label{fig:bsmm}
\end{figure}

As an example, Figure~\ref{fig:bsmm} shows the corresponding distribution of predicted values of $\bsmm$ at the best fit points. While the measurement of $\bsmm$ has been smeared according to a Gaussian distribution, this distribution clearly shows a non-Gaussian behavior. Its long tail shows that there is more space in the relevant CMSSM parameter space to go to larger values of $\bsmm$ than to lower values, which would be closer to the actual measurement. Nevertheless, the toy values are only in slight conflict with this measurement. 

\begin{figure}[t]
\includegraphics[width=0.5\textwidth]{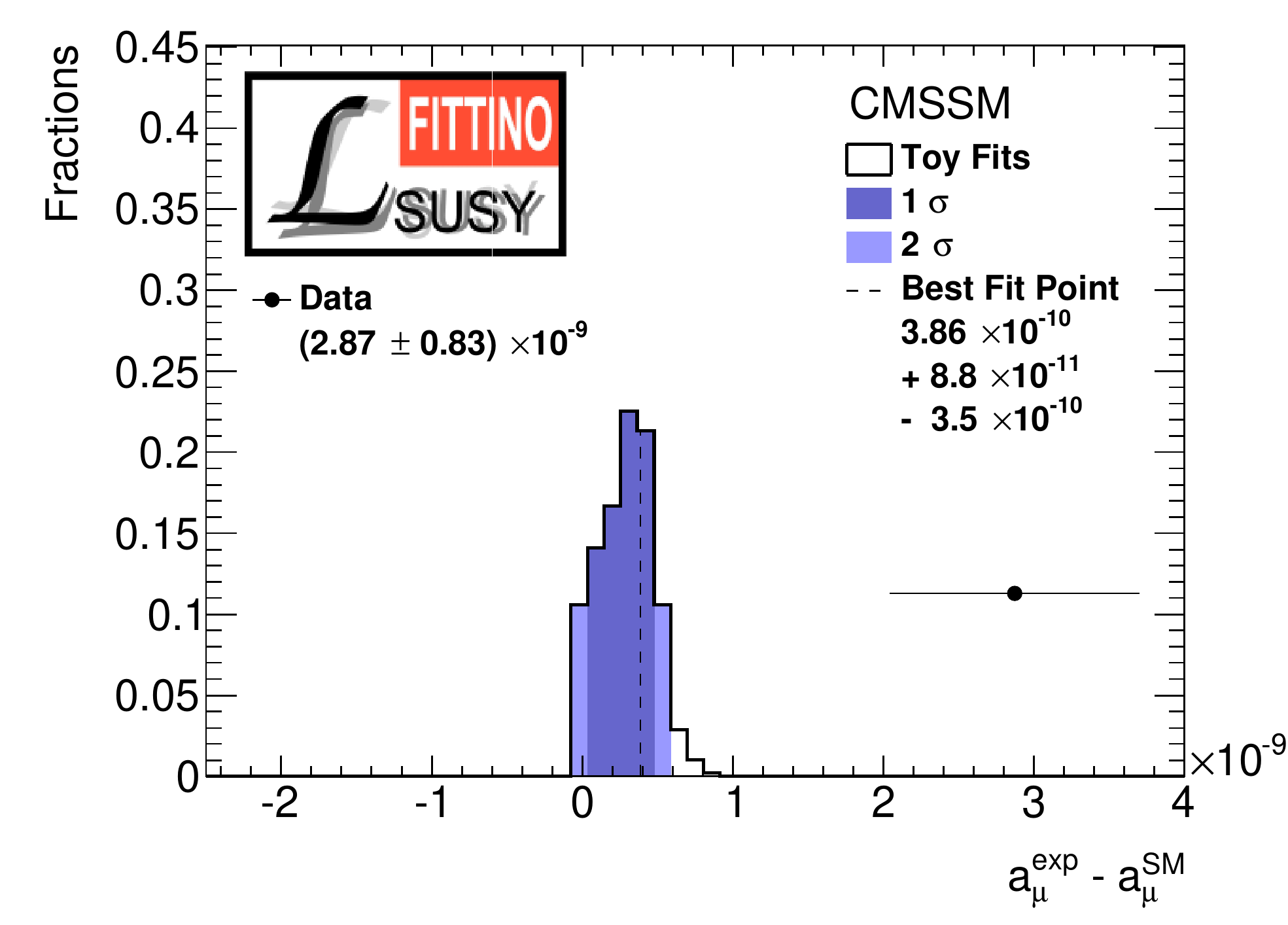}
\caption{ Distribution of predicted values of $a_\mu$ at the best fit points obtained from toy fits.}\label{fig:amu}
\end{figure}

The situation is different for $a_\mu$ as shown in Figure~\ref{fig:amu}. There is some variation in the best fit predictions of the toy fits, showing that in the relevant CMSSM parameter space $a_\mu$ still has some influence on the best fit point. However, this variation is small compared to the disagreement with the actual measurement. As a result, $a_\mu$ contributes to the $\chi^2$ much less in the toy fits than in the fit to the actual measurements.
Because of this, a minimal $\chi^2$ as bad as the one observed is found only in a small fraction of the toy fits.  

\begin{figure}[t]
\includegraphics[width=0.5\textwidth]{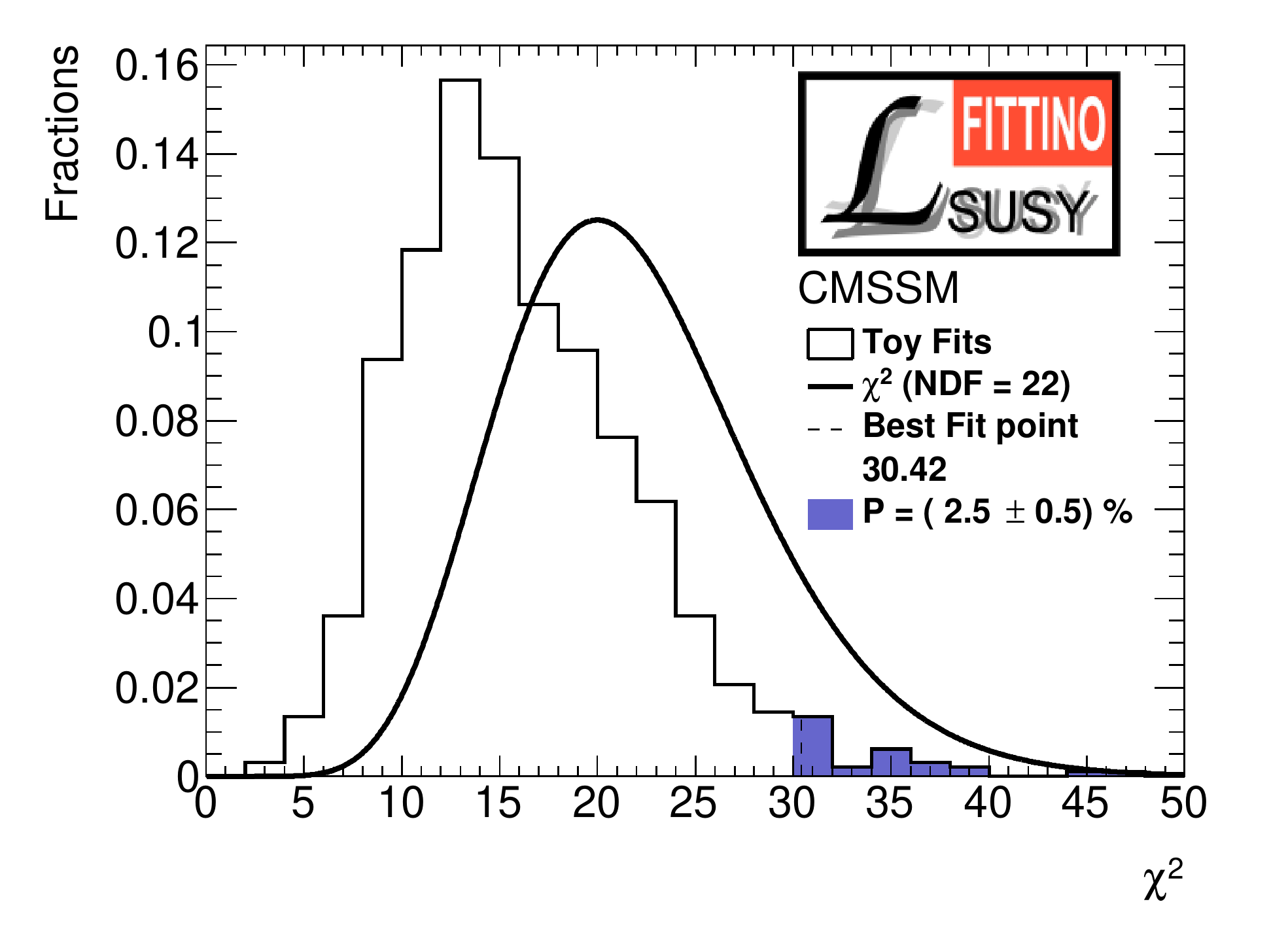}
\caption{ Distribution of minimal $\chi^2$-values obtained from toy fits. A $\chi^2$-distribution for 22 degrees of freedom is shown for comparison. }\label{fig:chi2}
\end{figure}

This can be seen in Figure~\ref{fig:chi2}, which shows the distribution of minimal $\chi^2$ values and in comparison a $\chi^2$-distribution for 22 degrees of freedom. Since the latter is shifted to the right, it overestimates the goodness of fit. Shown in blue is the fraction of toy fits giving a $\chi^2$ as bad as the one observed, resulting in a $p$-value of $(2.5\pm 0.5)\,\%$, excluding the CMSSM with $95\%$ confidence.

\section{Conclusion}\label{sec:conclusions} 
The first run of the LHC has improved our knowledge significantly: Among other results, there are the discovery of an SM-like Higgs boson, the first measurement of $\bsmm$, an improved top quark mass world average, and stronger constraints on new physics. In this work, all this information has been combined with the results from low energy and astrophysical experiments. In this way, we could show that the CMSSM as one of the simplest and most popular models is excluded with more than $95\%$ confidence. More details of this analysis will be given in an upcoming publication. 

Of course this is by no means a statement about supersymmetry in general. However, it sets the scene for applying the described method of global toy fits to more general models, where, for instance, the strong and electroweak sectors are disentangled. In this way, it will be possible to make full use of the potential of the upcoming results from the second run of the LHC.

\section*{Acknowledgments}
We are grateful to the organizers of the ICHEP 2014 conference for the opportunity to present this work. 
The work of MK was supported by the Deutsche Forschungsgemeinschaft through the 
collaborative research centre SFB-TR9 "Computational Particle Physics", by the German
Federal Ministry of Education and Research BMBF, and by the U.S. Department of Energy
under contract DE-AC02-76SF00515.
The work of TS was partially supported by a Feodor-Lynen research fellowship sponsored by the Alexander von Humboldt Foundation.



\nocite{*}
\bibliographystyle{elsarticle-num}
\bibliography{fittino}







\end{document}